\documentstyle[epsfig]{article}
\newcommand{\be}{\begin{eqnarray}}
\newcommand{\ee}{\end{eqnarray}}
\newcommand{\bra}[1]{\langle #1 |}
\newcommand{\ket}[1]{|#1\rangle}

\begin{document}

\title {
{\bf Toward the non-perturbative  description of high energy processes} \\
 E.V.~Shuryak\\
{\it Department of Physics and Astronomy, State University of New York,
     Stony Brook, NY 11794-3800} 
}
\date{\today}
\maketitle 
\begin{abstract}
General implications of existence of non-perturbative scales
and hadronic sub-structure for 
high energy processes are discussed. We propose that
the dependence of the cross section of $\bar q q$ dipoles
on their size d should deviate from $d^2$ when d becomes comparable to
substructure scale. 
Then we discuss
   Kharzeev-Levin pomeron model \cite{KL}, based on
ladder-type diagrams with $scalar$ resonances (scalar
$\pi\pi$ or $\sigma$ and the scalar glueball $G_0$). This channel
is truly unique, 
because instanton-induced attractive gg interaction \cite{Shu_82}
leads to 
unusually small sizes and strong coupling constants of these states,
supplemented by  unusually large mass scale, $M_0\approx 4 \, GeV$,  
  of the transition boundary to the perturbative regime.
As pomeron is a small-size object by itself, these resonances may play a
 special role in its dynamics. Furthermore,
we use more realistic
description of  the scalar gluonic spectral density  
without free parameters,
and slightly modify the model to get correct chiral limit.
 We conclude that the non-perturbative part of the scalar
contribution to the soft
pomeron intercept is  $\Delta  = .05\pm 0.015$, with comparable
contributions from both $\sigma$ and $G_0$. 

\end{abstract}
\vspace{0.1in}

1.Significant progress  of  non-perturbative QCD has been
mostly
related with approaches based on its Euclidean formulations:
lattice simulations, semi-classical theory based on instantons 
 etc. During the last decade we have learned a lot about
correlation functions and their spectral densities,
 hadronic wave functions and form-factors. Dramatically
different features of different
channels pointed out in \cite{NSVZ}  was confirmed and studied
in details \cite{SS_98}: below we
discuss one of the most striking cases, the
gluonic  $J^{PC}=O^{++}$ channel.

   However, little of this progress has so far contributed toward
 understanding of 
 high energy processes. True, it is difficult 
 to translate many of those tools into
Minkowski space. Elsewhere \cite{SZ} we will report some semi-classical
calculations aiming to bridge this gap, while we start
 this Letter with more general discussion of
  some qualitative ideas related  hadronic substructure
to high energy scattering.

  The most important lesson we learned is that the
  non-perturbative objects  in the QCD vacuum (and inside hadrons)
are not some shapeless soft fields, with
typical momenta of the order of $\Lambda_{QCD}\sim 1 fm^{-1}$, as it
was assumed in 70's. Instead we have semi-classical small-size
instantons
and very thin  QCD strings. The instanton
 radius peaks around the
 $\rho\sim 1/3 fm$ (\cite{Shu_82},for recent lattice data see
 \cite{anna}
and general review \cite{SS_98}).
 String energy (action) is concentrated
in a radius of .2 fm (.4 fm) in transverse directions \cite{Bali}.
 Both are small compared to typical hadronic size,
suggesting  a $substructure$ inside hadrons.
 
\begin{figure}[h]
\vskip 0.2in
\includegraphics[width=1.3in, angle=-90]{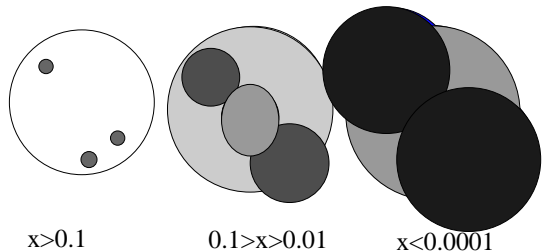}
\vskip 0.2in
\caption[]{
 \label{fig_snapshot}
A snapshot of the parton density in a nucleon at different x,
in an impact parameter plane.
}
\end{figure}

A snapshot of parton distribution in a
transverse plane inside the nucleon should look like 
indicated in Fig.(\ref{fig_snapshot}), for different x regions.
 These parton clusters originate from ``scars'' in the vacuum, being
perturbed
by  external
objects -- valence quarks and strings, and therefore
they must have the same transverse dimensions. One expects that these 
images of constituent quarks, diquarks and strings should be best seen  
at some intermediate x, before hadrons become black disks at very
small x (high energies).
 The non-perturbatively produced sea quarks supposed to be more concentrated
inside the constituent quarks\footnote{This concentration
should be enhanced for the polarized part of
the
sea. Sea quarks are found to be polarized
$opposite$
to  valence quark (and the nucleon), as the
instanton-based mechanism demands \cite{polarized}.
 }, while the string is supposed to be  gluonic.
A diquark cluster is also believed to be an instanton effect \cite{SS_98}. 
Finally, strong  $<\bar q q>$ modification inside the nucleon should
result in additional 
small density of sea quarks and gluons filling the whole disk (shown
by light grey in Fig.1). 

  If one prefers to use the language
of hadrons rather than fields, 
 existence of  two distinct
components can be viewed as being due to two different
scales for glueball and pion clouds, respectively.
However, using hadronic description in transverse plane
(or t-channel) is probably not very useful, because
such complicated and coherent field configurations as instantons and strings
can hardly be discussed well in this way.  

  These qualitative ideas were discussed in literature for long time.
Constituent quarks as clusters were discussed  e.g.in \cite{Shu_82},
and ``scars'' of two strings is behind Nachtmann-Dosch
model \cite{ND} 
of high energy hadron-hadron scattering. But they are still mostly
ignored by high energy practitioners, who think about partons as
being randomly distributed inside the hadronic disk.

 How can we tell whether such substructure really exists experimentally?  
The simplest process we have is
  deep-inelastic scattering (DIS). In the target frame, it can be viewed as
a scattering of   a dipole-like $\bar q q$ objects with variable size  $d\approx
 2/Q$, where Q is the
momentum transfer\footnote{This estimate works better for
  longitudinally polarized virtual photons, while for
transverse ones  large d tail is more significant, see \cite{FS}. }.
 The small-d dipoles 
measure only
the average gluonic field in the target\footnote{For exact 
definition  see \cite{Mueller}.}, but dipoles with d {\em comparable
to substructure scales} $\rho$ indicated above
should show a nontrivial behavior of
their cross section $\sigma(d)$ on d.
pQCD predicts that  $\sigma(d)/d^2$ is constant at small
d, but above some critical value $d> d_c\approx \rho$ we expect
this ratio to drop,
 because such large dipoles start to  miss the
 ``black spots'' in
the target.

  Phenomenologically,
 when HERA data were translated into such dipole cross section \cite{GBW,FS}
similar behavior
$\sigma(d)$ was indeed found, and at the right scale for d.
 In principle however, two different effects can be responsible for
 it. One, emphasized in  \cite{GBW,FS} especially
 for very small  $x$ is
 $saturation$ of the cross section, when the unitarity bound (or
 ``blackness'')
is reached.  Our idea suggest similar behavior of $\sigma(d)$,
 but even  at larger $x\sim 10^{-2}$ where the
spots are still ``grey'' and the  dipole cross section is not
 large enough to need shadowing corrections. In fact, the particular
 parameterization used in \cite{FS} have exactly this feature  for $all$
 x. 
 Further work is needed here to understand real magnitude of both
 effects,
due to shadowing and substructure, respectively.

   Another source of information about parton correlations in
   transverse
plane is diffraction. Clearly an inhomogeneous distribution
we advocate  enhances it, as compared to the homogeneous disk:
the blackness of spots in higher, and  there are more
edges at which diffraction may take place. 
Studies of how diffraction depends on $Q^2$
 at HERA is therefore very important. 
One possible parameterization is  hard-plus-soft pomerons
 \cite{2pomerons}, which places the boundary between soft and hard
 contributions
at $Q^2\sim 4-10 \, GeV^2$, or
$d \sim d_c$ we speak about.

 More generally, like it
happened for form-factors at similar $Q^2$  some time ago\footnote{
Although  $Q^2$ dependence of form-factors
 roughly follow  perturbative power
counting rules, their absolute magnitude significantly
 exceeds the pQCD
predictions. Non-perturbative approaches, such as
instanton-based calculation of the pion form-factor 
 \cite{BS}, are in quantitative agreement with data. 
}, after new round of works we may be forced
 to reassess where exactly
DIS is truly perturbative. The previous paradigm -- the
leading twist dominance down to Q as  low  
as $Q^2\sim 0.5 \, GeV^2$  -- seems now oversimplified.

   Hadron-hadron collisions are of course much more complicated.
Studies of inelastic
diffraction in $\pi p$,pp, pA and collision of two nuclei
\cite{fluct} definitely show
very large     O(1) fluctuations of the total nucleon cross section.
It is clear that it is completely incompatible with the picture
of a grey parton disk filled with multiple independent partons,
with dozens of degrees of freedom involved.  Only
{\it very few} degrees of freedom may drive those fluctuations.  
One obvious suspect is the distance between constituent quarks,
or the length of the string in Fig.1. Another is intermittent
blackness
of each constituent quark, presumably related to ``twinkling'' character of the
field strength distribution in the instanton vacuum.

    The pomeron itself is known to be a small-size object in transverse plane,
 as can be
  inferred from the fact that most of the t-dependence of pp scattering is explained
by nucleon form-factors, and also from
$\alpha'\sim (2 \, GeV)^{-2}$.
 It means when we see diffractive scattering of two nucleons,
those are related with diffraction of their small parts\footnote{Except 
at extremely very high energies, when the whole disk becomes black.}.

  The history of pomeron goes back 40 years:
  it still works very well\cite{DL}, new discoveries are being made
(such as existence of a pomeron polarization vector \cite{WA102}), 
but its microscopic dynamics still lacks
   theoretical understanding. pQCD promised to explain the
``hard pomeron'' \cite{BFKL}, and although recent calculation
\cite{next-to-lead} of the next-to-leading correction may put it in
doubt, hopefully there will be a way out 
\footnote{Similar things   happened before: let me just recall
  an optimistic example which may be unfamiliar to  many high energy
  physicists. Free energy for high temperature QCD has negative
     $ O(\alpha_s)$ correction of modest magnitude: but higher order
     ones show much larger corrections of different sign. However
     (unlike the pomeron intercept) one can calculate this free energy
non-perturbatively on the lattice. The result is about 15\% below free
gas, close to the $ O(\alpha_s)$ term. All
 high order corrections apparently canceled out! Furthermore, there are 
indications that some resummed or improved perturbation theory
exists, in which those cancellations happen explicitly.  }.
Non-perturbative approaches include  incarnations of the old 
multi-peripheral model (e.g. \cite{Bj}): but
they neither provide clear cut explanations of why  particular
hadrons should be used, nor give convincing quantitative predictions.
They also have no connection with Law-Nussinov 2-gluon exchange model,
which seems to be a very natural starting point explaining
 constant (not
growing with s) part of the cross section.

   New
 model for growing cross section and soft pomeron have been recently proposed
by Kharzeev and Levin  (KL) \cite{KL}. It includes
(i) a ladder made of two t-channel  longitudinal
gluons, as 
 in perturbative approach; 
(ii) while the s-channel ``rungs'' of the ladder
being replaced by production of  $scalar$ physical states. 
 Their main motivation  was  to apply some known
non-perturbative matrix elements, such as  gluon-to-
$\pi\pi$ transition near threshold. Furthermore, they 
inserted ``realistic'' scalar spectral density instead of
perturbative one below  some mass  $M_0$, suggested a
 schematic model for it and estimated a resulting 
 value for the  pomeron
intercept\footnote{ $\Delta$  enters the total hadronic cross section 
energy dependence as $\sigma(s) \sim s^\Delta$.} 
$\Delta\approx 0.08$, close to the phenomenological value \cite{DL}.
Their input was however very limited and therefore they have
treated gluonic scalar spectral density in a  very schematic way.
 It explains correctly qualitative features of the result (e.g.
its dependence on the number of colors and flavors $N_c,N_f$), but
 one may question the
accuracy of the resulting numbers.
Below we (i)  provide further
 motivation
for the KL approach, and also (ii)  
improve on their crude schematic model in several respects.
We explain why one should modify the expression  (\ref{delta})
for the pomeron intercept, and  get its meaningful
chiral limit. We use 
 other  information about scalar spectral density, with different
glueball parameters.
We found that
 quark-related (pion) contribution to  the  pomeron
intercept
$\Delta$ is reduced, and the glueball one is enhanced compared to KL numbers,
 making them comparable at the end.

2.To motivate the approach, let us start with the following 
question: How using  hadrons in a ladder diagram can
be consistent with the statements made above, that the pomeron
exchanges take
place between  {\it small parts} of the colliding hadrons,
such as constituent quarks or strings? 

Well, it depends: different hadrons have different sizes! 
For example,   multiple papers  use  chiral Lagrangians and
  pions propagating
$inside$  the nucleon\footnote{ Well known
 ultimate model of that kind was suggested by
Skyrme: in it a nucleon is made out of pions entirely.
As noted by Witten, large $N_c$ makes the
   nucleon static
 and pion field classical. Nevertheless,  the 
$R_N\Lambda_\chi >>1$ 
  condition is still needed to justify the model,
and because this parameter is
  not really large,  Skyrme model
  cannot be very accurate  at any $N_c$. }. It follows from chiral Lagrangian 
that it can be used     for momenta up to the so
called
chiral scale  $p < \Lambda_\chi\approx 1 \, GeV$.
Note:  it is the small pion size which matters here, not its mass. We will have
similar
situation for glueballs below.

With this in mind, we can move to the next question: Why is it
reasonable
to single out
scalar  $O^{++}$ gg channel? (Apart of the fact that we know few
related coupling constants.)
 Because both
 its prominent resonances --
the $\bar q q$ state $\sigma$
and the scalar glueball we call $G_0$ --
are very small-size.
 $\sigma$ is the pion's brother and its interaction is covered by the
 same chiral scale.
Remarkably,$G_0$  is even smaller, with
$R_{G_0}\approx .2 fm$ \cite{glueballsize,SS_95}. It is the smallest hadron we know,
setting a record of its kind.
It should be possible to construct some 
effective $G_0$ Lagrangian, applicable below some momentum scale $M_0$.
(For first attempts to build it see \cite{MigShi}, for discussion of the
magnitude
of $M_0$ see below.)
The qualitative reason why it is so compact is very strong 
instanton-induced attraction in the scalar channel due to small-size
instantons. As shown in \cite{Shu_82,SS_95}, in this case   diluteness
of the instanton vacuum $(\rho/R)^4\sim (1/3)^4$ (where $R \approx 1
\, fm$ is instanton mean separation)
is compensated by
classical enhancement factor $(8\pi^2/g^2(\rho))^2\sim 10^2$.  

Other channels do not have this feature. For example, another gg
channel one may think of is tensor $2^{++}$. However instanton field
do not have such component, and so they do not act in it. Consequently 
the tensor glueball has normal hadronic size,
 $R_{2++}\approx .8 fm$ according to \cite{glueballsize}, and it
 would be meaningless to consider its propagation 
 inside a nucleon. As experience with
 quark vector channels (where the situation is similar) shows, in such
 cases we have multi-hadron spectral density dual to perturbative one
down to  low scale $M_{2++}<<M_0$.

  KL have shown that the scalar  channel contribute the
following  ``non-perturbative part'' to the
pomeron intercept:
\be \label{delta}
\Delta \,\,=\,\frac{18\pi^2}{b^2}\,\int\,\frac{ d M^2}{M^6}
\,(\rho_{phys}( M^2 )-\rho^{pert}( M^2 ))
\ee
where $\rho_{phys}( M^2 ),\rho^{pert}( M^2 )$ are physical and
perturbative (gg cut)
spectral densities respectively.
The integrand in (\ref{delta}) is non-zero only for $M<M_0$  
 because at  $M>M_0$ pQCD works and two spectral
densities
become identical.

3.Let us  discuss  properties of scalar spectral
density
 $\rho_{phys}( M^2 )$, first in gluodynamics and then in QCD.

We use the same normalization of the
probing operator as KL
 \begin{equation}
\theta_\mu^\mu = 
\frac{\beta(g)}{2g} F^{a\alpha\beta} F_{\alpha\beta}^{a} \simeq 
- \frac{b g^2}{32 \pi^2} F^{a\alpha\beta} F_{\alpha\beta}^{a}; \label{trace} 
\end{equation}  so that 
 its perturbative spectral density is
\be
\rho_{pert}(M) = { \frac {
1}{4096}} \,{ \frac {b^{2}\,g^{4}\,({N_c}^{2}
 - 1)\,{M}^{4}}{\pi^{6}}} 
\ee
The spectral density should obey the low energy theorem
\cite{NSVZ}
\be
\label{let}
\int \frac{dM2}{M^2}
[\rho_{\rm phys}(M^2)-\rho_{\rm pert}(M^2)]
=-4\ \bra{0}\theta_\mu^\mu(0)\ket{0}
\ee
As alsomphasized in \cite{KL}, this relation
has historically provided  the first indication
  that there should be  large
non-perturbative
scale in scalar gg channel. 
Later more direct and quantitative assessment of this effect
  was proposed \cite{Shu_82}, based on small-size  instantons.

   In  gluodynamics $\rho(M)$ is dominated by 
 the contribution of the scalar glueball $G_0$, the lightest (and
 therefore stable) particle of this theory. 
Although phenomenologically its assignment to the observed scalar resonances 
is confused by mixing with $\bar q q$ resonances and is
still under debate, both multiple lattice works
 and the instanton model \cite{SS_95} point toward
$M_{G_0}= 1.5-1.7 \, GeV$.
As mentioned already, even more important is its
 small size, which leads
to a remarkably large coupling constant to gg current\footnote{We remind the reader that 
the units in the gluodynamics is
  traditionally defined by setting the string tension to be the same
  as in QCD.}, which according to \cite{SS_95} is
\be
\lambda_0 = <0|g^2 G_{\mu\nu}^2|G_0>\approx 16.\pm 2 \,\,\, GeV^3
 \ee
Substituting glueball contribution to 
the spectral density  
$
\rho_{G_0}=(b/(32 \pi^2))^2 \lambda_0^2 \delta(M^2-M_{G_0}^2)
$
one finds  the following contribution to the pomeron intercept
\be
\Delta_{G_0} \approx .03
\ee
This is not yet the complete non-perturbative part:  one
 still has to subtract the ``missing'' perturbative contribution for
$M<M_0$. 

As in   \cite{NSVZ}, we  determine $M_0$ from
a ``duality'' sum rule\footnote{Rather than from the low energy
  theorem (\ref{let}). With the simple sharp cutoff we use one cannot satisfy both, so
  we select duality because it closer to the integral we ultimately
  need. However the correlator calculated in \cite{SS_95} is in exact
  agreement
with this theorem.
 }
\be 
\int^{M_0}(\rho_{G_0}-\rho_{pert}){dM^2 \over M^4} =0
\ee
which ensures that correlator at small distances is not changed by
changing from $\rho_{pert}$ to  $\rho_{phys}$. Solving it for $M_0$,
one gets\footnote{In KL paper significantly smaller number $M_0\approx
  2.2\, \, GeV$ was used for QCD. It is dual to only $\sigma$ meson
  contribution, without a glueball. } 
$M_0\approx 4 /, \, GeV$. For a
boundary between hadronic and partonic descriptions
 it is unexpectedly high scale indeed.

The resulting
  contribution of ``missing perturbative states'' below $M_0$ leads to 
 negative contribution to $\Delta$ of about -.01.
 In total, we got $\Delta_{gluodynamics} \approx .02$,
 about twice the value estimated by KL.

\begin{figure}[t]
\vskip 0.4in
\includegraphics[width=1.3in,angle=-90]{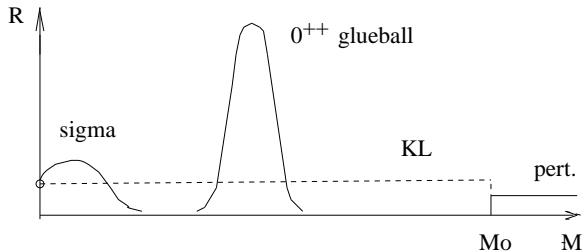}
\caption[]{
 \label{fig_sketch}
Schematic representation of the spectral density, as
$R=\rho(M)/\rho_{pert}(M)$ versus the total mass 
M. The small circle at M=0 is the chiral
Lagrangian prediction, the dashed line is the KL model, the solid line
is our spectral density, which includes the contribution of the glueball,
sigma meson and hadrons dual to perturbative gluons at large M.
}
\end{figure}

4.Now we return to the real world with light quarks, and consider the
sigma (or $\pi\pi$) contribution.
The major input of the KL paper is  the $\pi\pi$ coupling 
 at small M. It follows from the scale anomaly for chiral Lagrangian
 \cite{VZ} 
\be
\rho_{\pi\pi}^{M\rightarrow 0} = { 3 M^4 \over 32 \pi^2} 
\ee
which is larger than $\rho_{pert}$ because there is no $g^2$ \cite{KL}. 
The KL schematic model  assumed
that  $\rho(M)=
\rho_{\pi\pi}(M)$ for $all$ $M<M_0$ (see the dashed line in  Fig.(\ref{fig_sketch})).
 But
  this assumption cannot be true for larger  masses, because
 the  pions are known to interact strongly in this channel, 
forming the famous scalar $\sigma$ resonance.

  Since we know its parameters, 
 low energy $\pi\pi$ contribution into the correlation function in question
can be easily reconstructed. Consider the isovector vector
($\rho$-meson) 
channel, for which the spectral density is well known from  $e+e-$ collisions and $\tau$
lepton decay. In this case the $\pi\pi$ contribution  at
the threshold is trivial: the pion coupling  is
just  the pion charge.  Due to
pion attractive interaction, the spectral density  grows from  threshold, till it reaches the peak - the $\rho$-meson.
Its magnitude can therefore be 
fixed from the ``vector dominance'', using the normalization to the M=0
point and known $\rho$ meson width.

 Similar ``sigma-dominance''
should work even better, because
this resonance is very wide $m_\sigma\sim \Gamma_\sigma$.
\be
\rho_\sigma={ {3\pi^{2}\,{M}^{4}} \over {32}}\,
{ {M_\sigma}^{4}\over ({M^2} - {M_\sigma}^{2})^{2} + {M^2}\,\Gamma^{2}_\sigma }
\ee
Naive integration from the threshold ($2\,m_\pi$) gives  large
contribution
to pomeron intercept
$\Delta_\sigma^{naive} \approx .09$,
where we have used
$M_\sigma=.6\, \, GeV, \Gamma_\sigma=.4\, \, GeV$. (Coincidentally it is  close to what
KS got in their paper without sigma resonance,
 integrating till
(their) $M_0$.)

  However, this large contribution cannot be correct, because it
rest heavily on smallness of $m_\pi$.
In the chiral limit, when quarks and pions  become massless,
there is no threshold at $2m_\pi$, and the KL
$\Delta$  simply diverges.

To resolve  this problem, we should know the distribution over
  the so called {\it intrinsic} parton
transverse momenta\footnote{Deduced e.g. from $p_t$ distribution
of Drell-Yan pairs, after  perturbative effects due to extra gluon
emission at large dilepton mass M are subtracted.}. 
If gluonic partons are indeed nothing else but
expansion of classical
field of an instanton, their
transverse momenta 
 are  related to  basic non-perturbative scale $\rho$, the
mean size of QCD instantons. 
If so,
 one gets  its right magnitude $<p_t^2>\approx
(0.6 \, GeV)^2$ ,
and also predicts strong cut-offs, $both$ at high and low momenta. The
former is because the instanton is a finite-size object,
leading to a form-factor $\sim exp(p_t \rho)$. The latter happens because
the instanton field $A_\mu^a\sim \eta^a_{\mu\nu} x_\nu$
has a vortex-like shape with changing sign, so that its  projection
to  constant (or long-wavelength) field vanishes\footnote{
In fact,  there are experimental indications from diffractive
dissociation cross section for transverse virtual photon
is dominated by large $p_t$, which probably implies that indeed
$xg(x,k_t)\sim k_t^2$ at small $k_t$, see details in
\cite{Levin_talk}.}.
Therefore, gg collisions  with small invariant mass are suppressed.  

 Referring  to quantitative calculation  
elsewhere \cite{SZ}, here we simply modify
the KL expression
of (\ref{delta}) with a logarithmic accuracy,
 introducing into the KL integral  over transverse momentum
$k_t^2$ of the exchanged gluons, 
$\int dk_t^2/(k_t^2+M^2)^2$, a  finite cutoff $k_t^{min}$.
Now one of the 
    factors $1/M^2$ in (\ref{delta}) is modified, and
 \be \label{delta2}
\Delta \,\,=\,\frac{18\pi^2}{b^2}\,\int\,{ d M^2 \over M^4}
\,{(\rho_{phys}( M^2 )-\rho^{pert}( M^2 ))\over (M^2+(k_t^{min})^2)}
\ee
The unphysical divergence in the chiral limit is no longer there.
Although sensitivity to the cutoff value is formally logarithmic, it still 
 has significant effect on the intercept.
 With this modification, and $(k_t^{min})^2$ for the $pair$ to be
 equal to single parton $<p_t^2>$ mentioned, we find 
the $\sigma$ (or two-pion)
contribution  to be 
\be
{\Delta_\sigma|_{realistic}} \approx .03
\ee
It is now comparable to the glueball contribution
discussed above. This outcome is in fact more  natural
than the KS numbers, 0.08 and 0.01 for  $\sigma, G_0$ 
effects, because they are
is $O(N_f^2/N_c^2)$ (for $N_f=2$) and O(1), respectively.

5.{\bf Summary and discussion.} Finally, combining all contributions together
we get our final estimate of the non-perturbative
contribution to pomeron intercept, resulting from
$O^{++}$
hadronic ladder with $M<M_0\approx 4  \, \, GeV$ to be
\be \Delta  = .05\pm 0.015 \ee  
now with our guessed  uncertainties.

   How meaningful is this result? 
One still has to add to it the pQCD contribution, which is
\cite{BFKL,next-to-lead} 
with the region $M<M_0$ in the scalar channel subtracted.
With current uncertainties,  we do not know its value, and some 
readers may be disappointed at this point.
However,
   the progress is not zero.
First of all, the contribution we discuss in non-perturbative
$O(\alpha_s^0)$,  enhanced by classical instanton
effects $O(\alpha_s^{-1})$ compared to perturbative result. 
Second,  it is sufficiently close to the phenomenological value
$\Delta= 0.08$. Third, it can be experimentally tested by itself.

  Let us briefly indicate how it can be done. The discussed model
  claims that the ladders made of scalar resonances explains most of
the growing part of the cross section, namely $\delta \sigma_NN\approx
\sigma_{NN}(s_0) *0.05\, log \,(s/s_0)$. It means quantitative  
 predictions about multiplicities
of  these scalar resonances $\sigma, G_0$, both in  inelastic collision
 and their double diffractive
production. Sigmas are wide and distorted by low $p_t$ cut we do not
undertsnad well: so they  
 are difficult to trace down. The glueball however shows up
as relatively narrow resonance $f_0(1500)$, 
 found  in the double
diffractive
production. It has  good experimental signature: large $\eta\eta$ and $\eta\eta'$
decay modes. So its growing production with energy
may even affect s-dependence of the
$\eta/\pi$ ratio. But the hottest thing to understand is
the azimuthal distribution of
nucleons in double diffractive production:
  in fact the WA102 data (discussed in details in
\cite{Close_etal}) show surprisingly
 different  distribution for ``glueball-type''
resonances (such as $f_0(1500)$ we discuss) from that for quark-based mesons.
 
{\bf Acknowledgements} The author thanks D.Kharzeev,E.Levin,
L.McLerran,M.Strikman and I.Zahed for helpful discussions.
The work is partly supported by
the US DOE grant No. DE-FG02-88ER40388.


\end{document}